\documentclass[twocolumn,,nofootinbib,preprintnumbers,superscriptaddress]{revtex4}

\usepackage[dvipdf,dvips,dvipdfmx]{graphicx}
\usepackage{amsmath}
\usepackage{amssymb}
\usepackage{float}
\usepackage{wrapfig}
\usepackage{bm}
\usepackage{color}
\catcode`\@=11
\def\lsim{\mathrel{\mathpalette\@versim<}}
\def\gsim{\mathrel{\mathpalette\@versim>}}
\def\@versim#1#2{\vcenter{\offinterlineskip
\ialign{$\m@th#1\hfil##\hfil$\crcr#2\crcr\sim\crcr } }}
\catcode`\@=12

\newcommand{\p}{\partial}

\newcommand{\al}[1]{\begin{align}#1\end{align}}
\newcommand{\bp}{\begin{pmatrix}}
\newcommand{\ep}{\end{pmatrix}}

\newcommand{\bs}[1]{\boldsymbol}

\newcommand{\fn}[1]{\!\left(#1\right)}

\begin{document}

\title{Gauge hierarchy problem in asymptotically safe gravity\\--the resurgence mechanism}

\author{Christof \surname{Wetterich}}
\affiliation{Institut f\"ur Theoretische Physik, Universit\"at Heidelberg, Philosophenweg 16, 69120 Heidelberg, Germany}

\author{Masatoshi \surname{Yamada}}
\affiliation{Institut f\"ur Theoretische Physik, Universit\"at Heidelberg, Philosophenweg 16, 69120 Heidelberg, Germany}

\begin{abstract}
The gauge hierarchy problem could find a solution within the scenario of asymptotic safety for quantum gravity.
We discuss a ``resurgence mechanism" where the running dimensionless coupling responsible for the Higgs scalar mass first decreases in the ultraviolet regime and subsequently increases in the infrared regime.
A gravity induced large anomalous dimension plays a crucial role for the required ``self-tuned criticality" in the ultraviolet regime beyond the Planck scale.
\end{abstract}
\maketitle
\section{Introduction}
\begin{figure*}
\includegraphics[width=170mm]{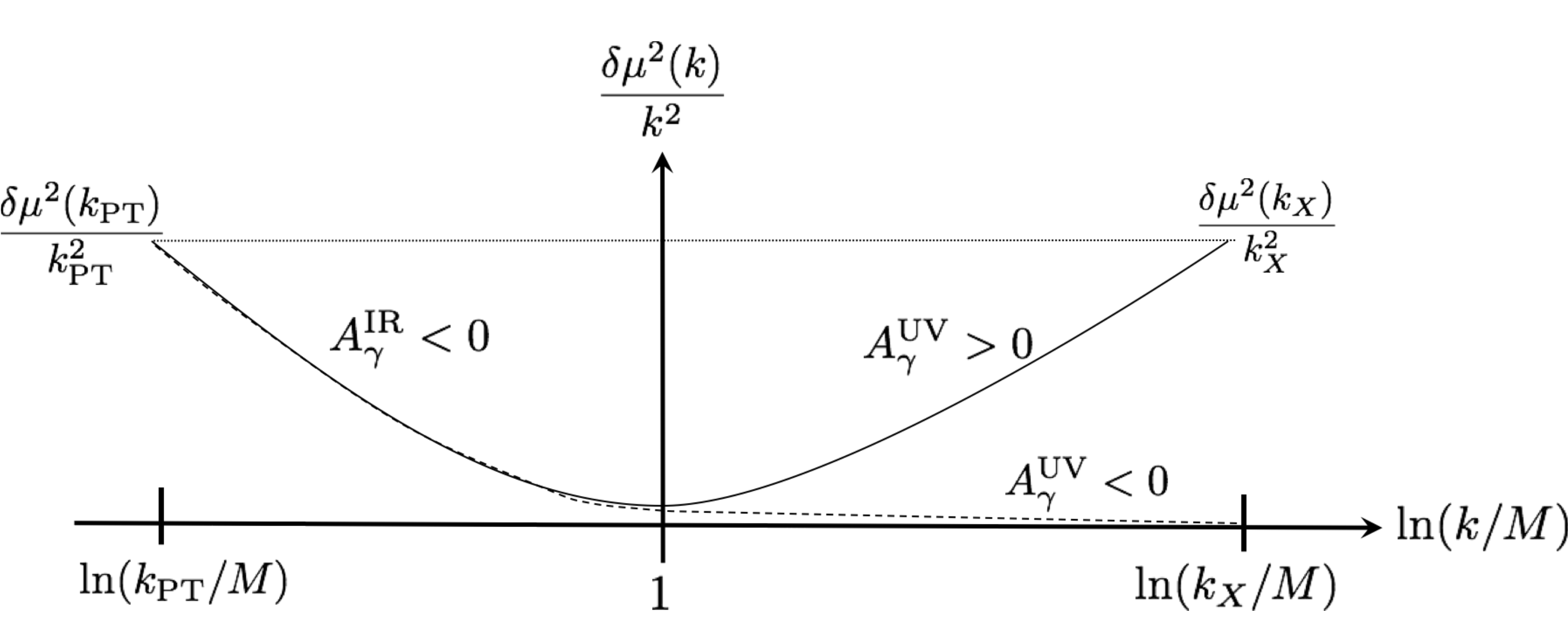}
\caption{The schematic figure of RG evolutions of the dimensionless coupling responsible for the mass of the Higgs scalar.
The solid line represents the RG running in the self-tuned criticality scenario.
The dotted line is the running with $A_\gamma<0$ for all energy regions.
The position of the $x$-axis corresponds to $\delta\mu^2/k^2=0$.
}
\label{massrg}
\end{figure*}
The discovery of the Higgs boson with a mass of 125 GeV at the Large Hadron Collider (LHC)~\cite{Aad:2012tfa,Chatrchyan:2012xdj},  together with the lack of discovery of ``new physics", indicates that the standard model (SM) of particle physics is valid at least up to the TeV scale.
From the observed value of the Higgs boson mass we learn that no obstacle prevents the validity of the SM up to very high scales, perhaps even the Planck scale.
The mystery deepens why the electroweak scale or Fermi scale is so much smaller than the Planck scale~\cite{Gildener:1976ai,Weinberg:1978ym}.
The tiny ratio between the Fermi scale and the Planck scale (or some scale of grand unification), called the gauge hierarchy, is a key mystery in elementary particle physics.
Attempts for its understanding, for example by supersymmetry, have generated many models that are tested at particle colliders.
The gauge hierarchy can be associated to a situation close to the zero temperature electroweak phase transition~\cite{Wetterich:1981ir}. 
The critical surface of an exact second order phase transition would correspond to a vanishing Fermi scale.
The understanding of the gauge hierarchy therefore answers to the question why the universe is so close to the phase transition.
In this letter, we argue that self-tuned criticality can be realized within asymptotically safe quantum gravity~\cite{Hawking:1979ig,Reuter:1996cp}.
An associated resurgence mechanism due to different anomalous dimensions in the ultraviolet (UV) and infrared (IR) regime can give a simple solution of the hierarchy problem.
\section{Flow in presence of a second order phase transition in the matter sector}\label{basic idea}
Let us sketch the situation of asymptotically safe gravity in the presence of a second order phase transition in the matter sector.
We denote by $\tau\fn{k}$ a characteristic dimensionless running coupling in the gravity sector, chosen 
such that for $\tau \gg1$ the gravity fluctuations decouple, while for $\tau \ll 1$ the gravity fluctuations are important.
As an example one may consider a running  (reduced) Planck mass $M\fn{k}$ and define
\al{
\tau =\frac{M^2\fn{k}}{k^2}=\frac{1}{8\pi G_N\fn{k}k^2},
}
such that $\tau\gg 1$ corresponds to a scale $k$ ``below the Planck mass", and $\tau\ll 1$ is beyond the Planck mass.
The precise definition of $\tau$ will not be important.
Asymptotic safety is realized if the flow of $\tau$ has a fixed point at some $\tau_*<1$,
\al{
k\frac{\p \tau}{\p k} =\p_t \tau &=\beta_\tau,&
\beta_\tau\fn{\tau_*}&=0,&
\beta_\tau\fn{\tau>\tau_*}&<0.& 
}
Due to $\beta_\tau\fn{\tau>\tau_*}<0$ the coupling $\tau$ is IR unstable.
For any ``initial value" $\tau>\tau_*$ it will increase as $k$ is lowered.
We only assume here that for the relevant solutions the coupling $\tau$ reaches one at some $k_\text{tr}$, and increases further for $k<k_\text{tr}$.

Let us next introduce the deviation $\delta \mu^2$ from the second order phase transition, typically a parameter in a scalar sector with dimension mass squared.
The flow of the dimensionless parameter
\al{
\gamma\fn{k}= \frac{\delta \mu^2\fn{k}}{k^2}
\label{dimensionless mass}
}
can never cross zero since a renormalization flow of couplings cannot cross the critical hypersurface which specifies a second order phase transition.
On the critical hypersurface scale symmetry becomes exact, such that $\gamma=0$ is a fixed point.
For small $\gamma$ the flow of $\gamma$ is typically linear in $\gamma$, with stability coefficient $A_\gamma$,
\al{
\p_t \gamma = A_\gamma \gamma.
\label{rgeq}
}
The stability coefficient will be influenced by the effective presence or absence of fluctuations in the gravitational sector.
This is reflected in the simplied, while qualitatively correct ansatz
\al{
A_\gamma=A^\text{IR}_\gamma\theta\fn{\tau-1} + A_\gamma^\text{UV} \theta\fn{1-\tau}.
\label{anomalous dim}
}

The UV fixed point of the combined gravity and matter system is located at $\gamma=0$, $\tau=\tau_*$.
The flow away from the fixed point depends critically on the sign of the UV-stability coefficient $A_\gamma^\text{UV}$.
For $A_\gamma^\text{UV}<0$ the coupling $\gamma$ is an IR unstable or relevant parameter. 
Any nonzero $\gamma$ will result in an increase of $|\gamma|$ as $k$ decreases.
In contrast, for positive $A_\gamma^\text{UV}>0$ the coupling $\gamma$ is IR stable or irrelevant.
Now the flow drives any deviation from the phase transition towards the critical surface at $\gamma=0$ as $k$ is lowered.  
This is called self-tuned criticality.
For positive $A_\gamma^\text{UV}$ the model realizes two types of scale symmetry.
Exact scale symmetry of the combined gravity and matter sector is realized at the UV-fixed point $\gamma=0$, $\tau=\tau_*$.
A second effective ``low-energy scale symmetry" in the particle sector, without  gravity, is realized for $\gamma=0$ and arbitrary $\tau$.
This second ``matter scale symmetry" is purely due to the second order character of the phase transition, independent of gravity.
It remains a good approximation as long  as $\gamma \ll 1$.
In particular, this can cover a large region $\tau\gg 1$, realizing approximate scale symmetry in the matter sector far below the Planck mass.

For $A_\gamma^\text{UV}>0$ the flow of $\gamma$ towards the phase transition surface continues only as long as $\tau<1$.
After the decoupling of the gravitational fluctuations the stability coefficient $A_\gamma$ typically turns negative, $A_\gamma^\text{IR}<0$.
In the SM, with small dimensionless couplings in the matter sector, one has
\al{
A_\gamma^\text{IR}=-2+A,
\label{dimless anomalous}
}
where the ``anomalous dimension" $A$ is a small, perturbatively calculable quantity.
Indeed, in the SM the anomalous dimension at perturbative one-loop level reads
\al{
A=\frac{1}{16\pi^2}\left( 2\lambda +6y_t^2 -\frac{9}{2}g^2-\frac{3}{2}g'{}^2 \right),
\label{anomalous dimension in perturbation}
}
where $\lambda$, $y_t$, $g$ and $g'$ are the quartic coupling constant of Higgs, top-Yukawa coupling constant, SU(2) gauge coupling constant and U(1) gauge coupling constant, respectively. 
From the observed masses, we have $A\approx 0.027$.

The solution of eqs.~\eqref{rgeq} and \eqref{anomalous dim}, with initial value set at some $k_X$, is straightforward.
The coupling $\gamma\fn{k}$ decreases in the range $k_\text{tr}<k<k_X$, and then increases for $k<k_\text{tr}$, cf.~Fig.~\ref{massrg}.
We call this ``resurgence".
Explicitly, one has
\al{
\gamma\fn{k}=\left( \frac{k}{k_\text{tr}}\right)^{A_\gamma^\text{IR}}\left( \frac{k_\text{tr}}{k_X}\right)^{A_\gamma^\text{UV}}\gamma\fn{k_X}.
\label{mass result}
}
A characteristic scale $k_\text{PT}$ for a substantial deviation from the flow on the critical hypersurface associated to the phase transition can be defined by $\gamma\fn{k_\text{PT}}=1$.
The physical scalar mass in the symmetric phase ($\gamma>0$) or phase with spontaneous breaking ($\gamma<0$) is given approximately by
\al{
m_\text{ph}^2=\gamma\fn{k_\text{PT}}k_\text{PT}^2=k_\text{PT}^2.
}

The resurgence mechanism can explain a natural gauge hierarchy.
Let us start at some scale $k_X$ without any very small couplings and take $\gamma\fn{k_X}=1$.
Associating $k_\text{tr}$ with the physical Planck mass $M=M\fn{k=0}$ yields
\al{
\frac{m_\text{ph}}{M}
&=\left( \frac{k_X}{M} \right)^{A_\gamma^\text{UV}/A_\gamma^\text{IR}}
\approx \left( \frac{k_X}{M} \right)^{-A_\gamma^\text{UV}/2} 
\approx 10^{-16}.
\label{ratio of masses}
}
In the last line we use $A_\gamma^\text{IR}=-2$ and associate $m_\text{ph}$ with the known mass of the Higgs scalar.
For any $A_\gamma^\text{UV}>0$ one can realize the observed hierarchy by starting at a suitable scale $k_X$ with a dimensionless coupling $\gamma\fn{k_X}$ of the order one.
Note that the dependence of $m_\text{ph}$ on the parameter $\gamma$ obeys $\p \ln\,m_\text{ph}/\p \gamma=\gamma/2$ and is order one.
Hence, it does not involve any fine-tuning.

In the SM the electroweak phase transition is second order up to small corrections to be discussed below.
The crucial role of the second order phase transition for the gauge hierarchy problem and the importance of the general form \eqref{rgeq}  of the flow equation have been emphasized in ref.~\cite{Wetterich:1981ir}.
First attempts to realize self-tuned criticality or the resurgence mechanism by obtaining $A^\text{UV}_\gamma>0$ from strong couplings of a fourth generation~\cite{Wetterich:1981ir} or strong Yukawa couplings~\cite{Bornholdt:1992up,Gies:2013pma} have not been successful.
In this letter we argue that asymptotically safe gravity offers a central new ingredient for realizing the resurgence mechanism.

\section{Irrelevant scalar mass in asymptotically safe gravity}
Evidences for asymptotically safe quantum gravity have accumulated from various approaches~\cite{Reuter:1996cp,Ambjorn:2004qm,Laiho:2011ya}; see reviews~\cite{Niedermaier:2006wt,Niedermaier:2006ns,Reuter:2012id,Codello:2008vh,Ambjorn:2009ts}.
Within a few GeV uncertainty the Higgs scalar mass $126\,\text{GeV}$ has been predicted for models of asymptotically safe quantum gravity~\cite{Shaposhnikov:2009pv}.
These results not only encourage the asymptotic safety scenario for quantum gravity but also give hints for physics at the Planck scale~\cite{Wetterich:2011aa} and cosmology~\cite{Wetterich:2014gaa}.

For our purpose the crucial quantities are the critical exponents for small deviations from the UV fixed point that governs asymptotic safety.
The stability coefficient $A_\gamma^\text{UV}$ is one of these exponents.
Let us consider a theory space spanned by a set of dimensionless couplings $g_i$.
The beta functions for the couplings are $\p_t g_i =\beta_i\fn{g}$.
It is essential for the asymptotic safety scenario that an UV fixed point exists for $g_i^*\neq 0$ at which $\beta_i\fn{g_*}=0$.
The flow of couplings around the UV fixed point is governed by the linearized equations
\al{
\p_t g_i \simeq \frac{\p \beta_i}{\p g_j}\bigg|_{g=g_*}(g_j-g_j{}_*)
=-T_{ij}(g_j-g_j{}_*).
}
The eigenvalues of the matrix $T$ are the critical exponents denoted by $\theta_i$.
Eigenvectors in coupling constant space with positive critical exponent correspond to relevant parameters.
In these directions the deviations from the UV-fixed point grow during the flow towards the IR.

The beta function has typically the following structure
\al{
\beta_i\fn{g} = -d_{g_i}g_i +f_i\fn{g},
} 
where $d_{g_i}$ is the canonical dimension of the coupling $g_i$ and $f_i\fn{g}$ is the contribution from fluctuations.
When the off-diagonal part of $\frac{\p f_i}{\p g_j}|_{g=g_*}$ is negligible, the critical exponent can be approximated by
\al{
\theta_i \simeq  d_{g_i} - \frac{\p f_i}{\p g_i}\bigg|_{g=g_*}.
}
The contribution $ \frac{\p f_i}{\p g_i}|_{g=g_*}$ is called the anomalous dimension around the UV fixed point.

In our context the stability coefficient $A_\gamma^\text{UV}$ is the negative of a critical exponent, namely the one describing the deviation from a phase transition in the matter sector.
The UV-anomalous dimension is $A_\gamma^\text{UV}+2$, while the cofficient $A$ in eq.~\eqref{dimless anomalous} corresponds to the anomalous dimension in the IR-regime.
The functional renormalization group (FRG)~\cite{Wilson:1973jj,Polchinski:1983gv,Wetterich:1992yh} is the convenient tool for the computation of fixed points and critical exponents.
We employ here the formulations based on the effective average action~\cite{Wetterich:1992yh} which has proven to achieve reliable results for many systems where couplings are large and perturbation theory fails.

We are interested in models of gravity coupled to a scalar field with a second order phase transition, for example due to spontaneous symmetry breaking of a discrete $Z_2$-symmetry.
The coupling $\gamma$ is to a good approximation given by $\delta \mu^2/k^2$, with $\delta\mu^2$ a scalar mass term.
For such systems, also coupled to additional fermions, first FRG studies have been performed in the papers~\cite{Percacci:2003jz,Narain:2009fy,Oda:2015sma,Eichhorn:2016esv,Henz:2016aoh}.
Even though the truncations employed so far only involve a few couplings these first results shed light on the possible realization of the resurgence mechanism.

In the scalar-gravity system without fermions the stability coefficient (negative of the critical exponent for the dimensionless scalar mass term) was obtained as $A_\gamma^\text{UV}\approx -0.117$~\cite{Percacci:2003jz,Oda:2015sma}.
The gravity induced anomalous dimension is large, but not enough to make the resurgence mechanism work.
In contrast, for the Higgs-Yukawa model non-minimally coupled to gravity the scalar mass term has been observed to become an irrelevant coupling~\cite{Oda:2015sma}.
The fermionic fluctuations influence the location of the fixed point in the gravitational sector.
In turn, this affects the critical exponents and the first results for the system indicate a positive stability coefficient $A_\gamma^\text{UV}\approx 0.5$~\cite{Oda:2015sma}.
In ref.~\cite{Oda:2015sma} the authors have studied a truncated theory space spanned by six couplings, namely the Planck mass, the cosmological constant, the non-minimal scalar-gravity coupling, the scalar mass and quartic coupling, and the Yukawa coupling.
If the positive value of $A_\gamma^\text{UV}$ is confirmed in extended truncations the resurgence mechanism can work above the Planck scale.
Using the relation \eqref{ratio of masses} for $A_\gamma^\text{UV}=0.5$, one finds $k_X\approx 10^{82}\,\text{GeV}$.

\section{Quadratic divergence , ``fine-tuning problem" and intermediate scales}
In view of the extended discussion on naturalness and fine-tuning problems the reader may ask if the resurgence mechanism can work in the SM. 
Let us emphasize from the outset that the scale symmetry and the ``loss of memory" at fixed points typically introduce relations that may not look obvious from the point of view of too naive perturbative order of magnitude estimates.

We begin with the fine-tuning problem.
In a simple one loop calculation the Higgs mass-squared is given as
\al{
\mu^2=\mu_0^2 + c\Lambda^2,
\label{fine tuning mass}
}
where $\mu_0$ is the bare Higgs mass, $c$ is a factor depending on different couplings of particles to the Higgs boson and $\Lambda$ is some effective UV-cutoff, typically much larger than the Fermi scale, $\Lambda\gg k_\text{PT}$.
If $\Lambda=M$, a renormalized mass term $\sim (100\, \text{GeV})^2$ requires a fine-tuning of 32 digits between the bare mass $\mu_0^2$ and the one-loop effect $c\Lambda^2$.
This is one aspect of the so-called fine-tuning problem. 

The bare mass term parametrizes the position of the critical hypersurface corresponding to the electroweak (zero--temperature) phase transition~\cite{Wetterich:1983bi,Aoki:2012xs}.
While the existence of the phase transition is generic, the precise location of the critical surface in coupling constant space will depend on the detailed model, the parametrization of the model and regularization procedures.
In fact, already in the one loop computation the factor $c$ strongly depends on the regularization scheme.
For Pauli-Villars type regularizations $c$ explicitly depends on details such as the regulator mass.
In particular, dimensional regularization gives $c=0$.
The fine-tuning problem concerns the computation of the precise location of the critical hypersurface in some particular loop expansion scheme.
This is not a quantity of interest.

In contrast, the deviation $\delta \mu^2$ from the critical hypersurface does not depend on the regularization scheme.   It is universal and physically meaningful parameter.
In the context of flow equations $\delta \mu^2\fn{k}$ involves the running scalar mass term $\mu^2\fn{k}$~\cite{Wetterich:1990an}
\al{
\delta \mu^2\fn{k}=\mu^2\fn{k} - {\mu}_*^2\fn{k}.
}
Here $ {\mu}_*{}^2\fn{k}$ is the critical trajectory or scaling solution, corresponding to the flow of the critical hypersurface in theory space.
For a second order phase transition the scaling solution obeys $\mu^2_*\fn{k\to 0}\to 0$.
The evolution of $\delta \mu^2\fn{k}$ is described by eqs.~\eqref{dimensionless mass} and \eqref{rgeq}.
The IR-anomalous dimension $A$ in eq.~\eqref{dimless anomalous} is a universal quantity that can be computed in perturbation theory without any fine tuning.
No fine-tuning problem exists from the point of view of the renormalization group~\cite{Wetterich:1981ir,Wetterich:1987az,Wetterich:1990an,Wetterich:1991be}.
Technical naturalness of a Fermi scale much smaller than $M$ can be related to an additional symmetry at a second order phase transition, namely scale symmetry~\cite{Wetterich:1987az}.

Other possible objections to the resurgence mechanism concern the the possible presence of intermediate scales, as for example in grand unification (GUT). 
The term $c\Lambda^2$ in eq.~\eqref{fine tuning mass} is now replaced by $\lambda_G\langle \varphi_G \rangle^2$, where $\langle \varphi_G \rangle$ is the expectation value characterizing spontaneous symmetry breaking of the GUT-symmetry, and $\lambda_G$ is some coupling constant --e.g. a combination of quartic scalar couplings.
For $\langle \varphi_G \rangle$ much larger than the Fermi scale the fine-tuning problem seems to reappear.
The issue is, however, precisely the same as for the quadratic divergence.
It concerns the precise location of the critical hypersurface, not is existence and the flow of $\delta\mu^2$~\cite{Wetterich:2011aa}.
(It will affect the precise form of the scaling solution $\mu^2_*\fn{k}$ for $k^2 \gtrsim \langle \varphi_G \rangle$.)

Discarding the rather irrelevant issue of the precise location of the critical surface in coupling constant space the core of the gauge hierarchy problem remains the question why the Fermi scale is so much smaller than the Planck scale.
In other words, why is the world so close to criticality?
The resurgence mechanism provides for a possible answer.

\section{Fermi scale in the resurgence mechanism}
To conclude, we ask the question which physics finally determines the value of the Fermi scale or $k_\text{PT}$.
There are possible high energy and low energy answers.
If the flow beyond the Planck mass continues without further changes towards the UV-fixed point, the scale $k_X$ for setting initial values can be taken to infinity.
For an exact second order phase transition eq.~\eqref{mass result} yields $k_\text{PT}=0$ for $k_X\to \infty$ and any finite $\gamma_X$.
An infinitely long attraction towards the phase transition in the UV regime cannot be compensated by any finite ``duration of repulsion" in the IR-regime.
The $k$-flow of $\gamma$ has an analogy in cosmology with the time dependence of the deviation from the critical density.
The phase transition at $\gamma=0$ corresponds to the fixed point in the cosmological evolution, $\Omega-1=0$.
The period of attraction to the fixed point at $\Omega=1$ corresponds to inflation, the repulsion period to matter and radiation domination.
For an infinite number of $e$-foldings in inflation the present value would be exactly at $\Omega=1$, and this would remain so in any finite future even if the repulsion would continue (not realized in presence of dark energy).

The limit $k_X\to \infty$ implies ``classical" scale invariance of the low energy theory.
One requires a low energy explanation why the Fermi scale does not vanish.
The basic reason for this possibility is the fact that the electroweak phase transition is only approximately second order.
Running dimensionless couplings as the gauge couplings turn it to a weakly first order transition for which scale symmetry is not exact.
For the observed values of couplings in the SM the dominant effect comes from the running of the strong gauge coupling.
By chiral symmetry breaking this gives a lower bound for the Fermi scale in the range around $100\,\text{MeV}$.
Other effects from running couplings in the SM appear to be even weaker.
A low energy determination of the Fermi scale seems to require on extension of the SM, perhaps with a possible Coleman--Weinberg mechanism~\cite{Englert:2013gz}.
Such extended low energy models could be tested by forthcoming experiments at particle colliders.

The high energy determination of $k_\text{PT}$ becomes possible if the UV-flow with $A_\gamma^\text{UV}>0$ does not continue towards $k\to \infty$.
A new regime may finally describe the approach to an UV fixed point for $k>k_X$.
In this case the initial value $k_X$ for which $A_\gamma^\text{UV}>0$ holds can be finite and a determination of $k_\text{PT}$ by an analogue of eq.~\eqref{ratio of masses} becomes possible.
In our analogy to inflation this corresponds to a duration of inflation for a finite number of $e$-foldings.

Independently of the precise mechanism which determines the value of the Fermi scale the resurgence mechanism offers a natural explanation for a huge gauge hierarchy.
Forthcoming quantum gravity computations will reveal if the required positive value of $A_\gamma^\text{UV}$ is indeed realized.

\subsection*{Acknowledgements}
M.Y. thanks Kin-ya Oda for helpful discussion.
This work is supported by the DFG Collaborative Research Centre SFB 1225 (ISOQUANT) and by ERC Advanced grant ERC-2011-ADG 290623 (FUNREN).

\end{document}